# Design and implementation of a high-density sub-nanosecond timing system for a C-band photocathode electron gun test platform

**Peng Zhu** [a,b], **Kangjia Xue**[a,b], **Lin Wang**[a,b,c], **Yuliang Zhang**[a,b], **Yongcheng He**[a,b], **Xuan Wu** [a,b], **Mingtao Li** [a,b,c], **Sinong Cheng** [a,b], **Xiaohan Lu** [a,b,c], **Shiming Jiang** [a,b,c], **Xiao Li** [a,b,c,*]

[a] *Institute of High Energy Physics, Chinese Academy of Sciences,
  Beijing 100049, China*

[b] *Spallation Neutron Source Science Center,
  Dongguan 523803, China*

[c] *University of Chinese Academy of Sciences,
  Beijing 100049, China*
  *E-mail*: `lixiao@ihep.ac.cn`

ABSTRACT: This paper presents the design and implementation of a high-density, deterministic trigger distribution system tailored for the C-band photocathode electron gun test platform at the Southern Advanced Photon Source (SAPS). Implemented within a scalable 6U VME modular architecture, the system achieves high-density integration by consolidating a master controller, clock distribution network, and 80 heterogeneous output channels into a single chassis. This design leverages a high-performance FPGA core combined with custom backplane interconnections to establish a master-slave topology, significantly reducing the system footprint compared to stacked standalone generators. To guarantee timing determinism in high-noise environments, precise placement and timing constraints are applied to the FPGA logic, while optical isolation is employed to mitigate electromagnetic interference. Furthermore, a dual-channel SFP optical signaling architecture enables seamless expansion to 160 synchronized channels. A remote control framework based on a serial server and a virtual machine Input/Output Controller (IOC) facilitates flexible configuration. Performance tests demonstrate adjustable trigger frequencies from 1 Hz to 100 Hz, with delays and pulse widths tunable from 0 to 10 ms at a resolution of 10 ns (or the RF period). The local electrical output exhibits an ultra-low RMS jitter of 6.55 ps (60 ps peak-to-peak). For remote optical distribution, the system maintains a sub-nanosecond RMS jitter of 119.5 ps, with peak-to-peak variation confined to 1 ns due to the combined effects of transceiver optoelectronic conversion (utilizing HFBR-1414T/2412T modules) and fiber transmission. The system has been successfully commissioned and is currently in reliable routine operation, verifying the architecture as a robust, highly integrated, and cost-effective solution for compact accelerator facilities.

**Contents**

## 1. Introduction

The Southern Advanced Photon Source (SAPS) represents a significant leap in the development of next-generation diffraction-limited storage ring light sources[1]. In its full-energy injection scheme, a photocathode electron gun is employed as the primary electron source. To validate the critical technologies associated with this injector, a dedicated photocathode gun test platform has been established[2]. Central to the reliable operation of this platform is the requirement for a timing system capable of triggering spatially distributed subsystems, including drive lasers, RF power sources, and beam diagnostics, with sub-nanosecond precision[3]. Consequently, a dedicated timing solution that effectively combines high channel density, cost-effective scalability, and robust synchronization stability is indispensable.

     Currently, high-density timing signal generation in accelerator facilities is typically implemented using three mainstream architectures[4]. The first and most prevalent is the Event Timing System (e.g., Micro-Research Finland[5]), which broadcasts event codes over optical links to receiver nodes housed in MicroTCA.4 or VME64x crates[6][7]. While offering excellent synchronization performance, these systems are often cost-intensive and complex to deploy for small-to-medium-scale test stands. The second approach utilizes the White Rabbit (WR) network, based on IEEE 1588-2008 and Synchronous Ethernet standards, which provides precise, long-distance, distributed synchronization[8][9]. However, achieving high-density electrical fan-out (e.g., 80 channels or more) within a compact footprint typically necessitates stacking multiple WR nodes, thereby increasing system complexity and integration overhead. The third method involves cascading commercial digital delay generators [10]. Although straightforward to implement, this approach suffers from accumulated jitter in daisy-chained configurations and lacks seamless integration with distributed control frameworks such as EPICS [11].

To address these limitations, particularly concerning scalable channel density and cost-efficiency, this paper presents the design and implementation of a high-precision clock generation and timing distribution system tailored for the C-band photocathode electron gun test platform. Commercial alternatives, such as the MRF event timing system and White Rabbit networks, provide state-of-the-art stability and long-distance deterministic synchronization, making them the gold standard for large-scale accelerator facilities. However, implementing these systems on a compact, medium-scale test platform introduces significant overhead. To generate an 80-channel localized fan-out using commercial event receivers, one must stack multiple interface boards and active crates, increasing hardware costs to nearly ten times those of a customized solution. Additionally, commercial timing systems often operate on rigid, event-driven protocols, making it exceptionally difficult and opaque to implement facility-specific, non-standard logic modifications at the firmware level. During the R&D commissioning and high-voltage conditioning phases, physicists often need to debug timing parameters iteratively and integrate bespoke logic rapidly, such as non-standard pulse-burst modes or hardware-level machine protection interlocks. Unlike general-purpose commercial alternatives[12] [13][14], the proposed system adopts a scalable 6U VME modular architecture centered on a high-performance FPGA core. A critical design feature is the use of custom backplane interconnections to establish a master–slave topology, effectively eliminating the timing uncertainty inherent in conventional shared-bus protocols. The system supports up to 80 independent output channels through interchangeable optical and electrical interface boards, providing exceptional flexibility for diverse instrumentation requirements.

To ensure timing determinism in high-noise environments, the logic design incorporates rigorous physical constraints, and extensive optical isolation is employed to mitigate electromagnetic interference. Furthermore, a control architecture based on a serial server and a virtual machine Input/Output Controller (IOC) enables remote visualization and flexible configuration. By leveraging the standard VME mechanical architecture and utilizing a custom passive backplane instead of active protocol switches, the design achieves an optimal balance between performance and cost. Additionally, rather than relying on dedicated host CPUs and hardware-specific kernel-space drivers, typically required for commercial VME/MTCA timing modules, this system utilizes a simple serial server and a virtual machine IOC. This hardware-OS decoupling provides a lightweight, OS-agnostic 'plug-and-play' control alternative that significantly reduces the deployment and maintenance overhead for small-scale test benches.

## 2. System overview and timing requirements

A comprehensive pre-research program has been established to validate the critical technologies for the SAPS, specifically the C-band high-gradient electron gun and traveling-wave accelerating structures. As depicted in figure 1, the test facility's commissioning is executed in two strategic phases, each imposing distinct requirements on the timing distribution system.

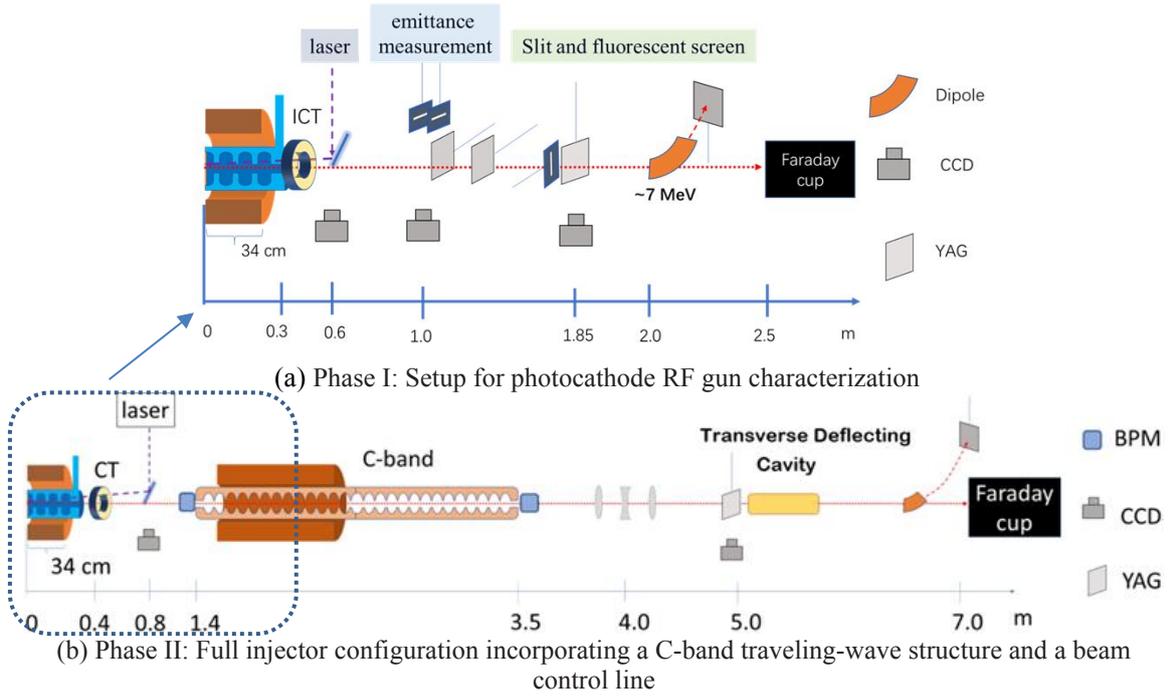

(a) Phase I: Setup for photocathode RF gun characterization

(b) Phase II: Full injector configuration incorporating a C-band traveling-wave structure and a beam control line

**Figure 1.** The schematic layout of the SAPS C-band photocathode electron gun test platform.

    To meet the stringent beam dynamics requirements of the C-band photocathode gun test platform, it must deliver synchronization triggers with sub-nanosecond determinism relative to the electron bunch arrival time and maintain sub-nanosecond jitter stability. This precision is crucial for preserving longitudinal phase space quality during acceleration. Within the FPGA fabric, digital frequency division and synchronous counter logic synthesize the specific repetition rates required by various subsystems. High-resolution counters with synchronous reset logic govern key pulse parameters, repetition rate, pulse width, and relative delay, enabling fine-grained, real-time adjustment of correlated timing channels. The system is designed to enhance beam commissioning efficiency by offering a wide dynamic range, supporting both coarse adjustments (e.g., 1 µs steps for width control) and ultra-fine delay tuning. The detailed technical specifications and adjustment ranges are summarized in Table 1.

**Table 1** Requirement of the global synchronization timing

| Time | Definition and Comment |
| --- | --- |
| Number of signals | >20 |
| Frequency setting | 1/5/10/25/50/100 Hz |
| RMS jitter of trigger signals | <1 ns |
| Delay range of trigger signals | 0 - 10 ms |
| Delay adjustment step of trigger signals | 10 ns / period of the reference RF signal |
| Pulse width range of trigger signals | 0 - 10 ms |
| Pulse width adjustment length of trigger signals | 10 ns / period of the reference RF signal |

## 3. Hardware architecture design and implementation

### 3.1 Architecture Design

As illustrated in figure 2, the timing system is based on a "one-master, multiple-slaves" star topology, comprising five key functional components: the central core logic board (CLB), output interface boards (OIBs), a custom VME J2 breakout backplane, and remote signal reception terminals. Unlike conventional VME systems that rely on shared-bus protocols for data transfer, the proposed architecture utilizes the standard 6U VME chassis principally for mechanical support and power distribution. The CLB, located in the central master slot (slot 4), serves as the primary timing engine, while the OIBs are placed in the available slave slots (slots 2, 3, 5, 6, and 7). All timing signals are distributed through a dedicated, custom-designed VME J2 breakout backplane. This modular design allows a flexible combination of optical and electrical interface boards within a single 16-channel chassis, facilitating rapid maintenance and reducing mean time to repair (MTTR) during accelerator commissioning. This high-density integration consolidates 80 synchronized channels into a single crate.

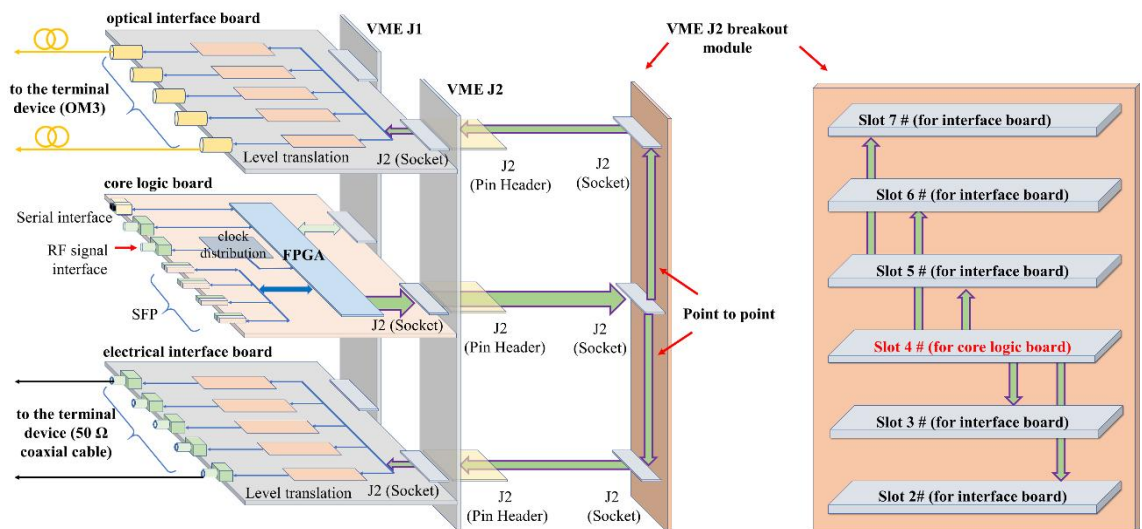

**Figure 2.** The schematic and function diagram of hardware architecture.

    In contrast to conventional high-channel-count solutions, which commonly rely on daisy-chaining standalone commercial delay generators, such as the SRS DG645, and accumulate additive trigger jitter, or depend on stacking network-based event receivers requiring active switching nodes, the proposed system utilizes a strict, physical-layer star topology within a single chassis. A significant innovation in this design is the repurposing of the VME J2 connector to create a custom, entirely passive hard-wired interconnect. The central core logic board directly distributes the reference clock and triggers to all output boards through dedicated point-to-point backplane traces. By bypassing active network switches and avoiding daisy-chained topologies, this parallel distribution method ensures that system-wide timing jitter is primarily influenced by the master FPGA's clocking network and active logic components in the distribution path, such as optical drivers, transimpedance amplifiers (TIAs), and receiving CPLDs. Additionally, the passive backplane effectively minimizes impedance mismatch and static inter-channel skew.

## 3.2 Core Logic Board Design

The core logic board functions as the central control and processing unit and is constructed around a high-performance Field-Programmable Gate Array (FPGA). The design employs a Xilinx Spartan-6 series FPGA, complemented by several essential subsystems, including a power supply module, a clock distribution module, a high-speed serial link interface, and a suite of expansion interfaces, including JTAG, serial interface and VME buses. These interfaces support a wide range of applications and facilitate future functional upgrades. The overall design framework and photograph is depicted in figure 3.

The core logic board supports flexible clocking schemes to ensure phase coherence. In external mode, it receives the accelerator's master reference frequency through a front-panel RF interface. An onboard clock distribution network conditions this signal, ensuring that the FPGA's internal logic counters are strictly phase-locked to the accelerator's RF period. For standalone operations, a high-precision 100 MHz reference is generated using an onboard low-jitter, temperature-compensated crystal oscillator[15]. This reference is buffered by a low-jitter fan-out chip to drive internal logic and optical transceivers. The FPGA is capable of synthesizing up to 80 independent trigger signals. To eliminate arbitration latency and maintain timing precision, these signals are routed exclusively through user-defined pins on the VME J2 backplane connector, thereby bypassing the shared VME bus. A serial communication interface facilitates real-time interaction with the high-level control system, permitting dynamic adjustments of delay, pulse width, and repetition mode without interrupting beam operations. Furthermore, high-precision trigger signals are transmitted via a high-speed optical transceiver connected to the FPGA's general I/O, while the remaining high-speed serial link modules interface with the FPGA's Gigabit Transceiver (GTP) modules, thereby enhancing operational flexibility across diverse experimental configurations.

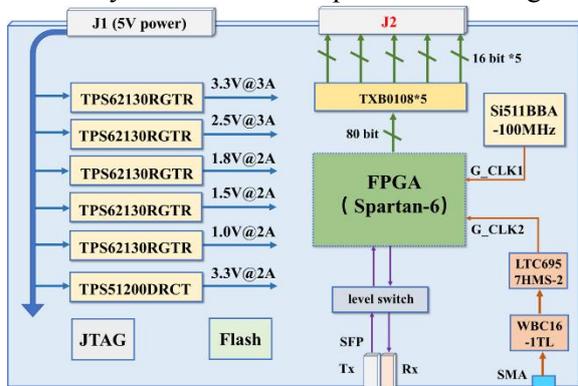

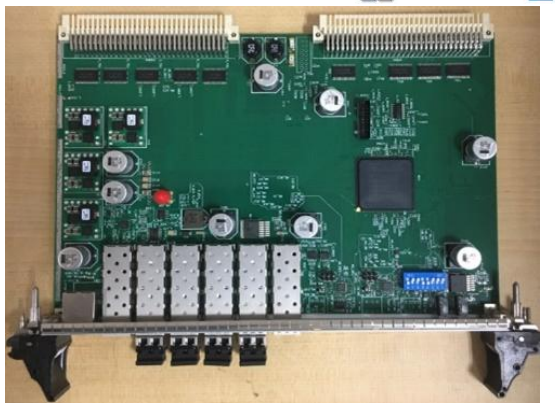

**Figure 3.** (a) Schematic diagram of the CLB. (b) Photograph of the hardware.

## 3.3 Output Interface Boards Design

To accommodate a variety of downstream instrumentation, up to five output interface boards can be installed in the slave slots simultaneously. Each board provides 16 output channels and conditions the logic signals received from the core logic board, as shown in figure 4(a). Two hardware variants have been developed:

- **optical interface board**: designed for long-distance transmission (> 20 m) or operation in high-electromagnetic-interference environments, as depicted in figure 4(b), this variant employs HFBR-1414T optical transceiver modules to convert electrical trigger signals into optical pulses. This implementation ensures complete galvanic isolation, effectively eliminating ground loops and suppressing common-mode noise.
- **electrical interface board**: intended for local diagnostics, such as connection to oscilloscopes or digitizers, as depicted in figure 4(c), this variant incorporates high-speed buffer-driven level-shifting circuitry. It delivers standard TTL-compatible signals with 50 Ω impedance matching through LEMO connectors, ensuring an achieved 10%–90% signal rise time of less than 1 ns and high signal integrity.

The interface boards incorporate only the level-shifting and driver circuitry essential for signal transmission, without any redundant components. This minimalistic design provides ample PCB real estate to implement strictly isochronous routing across all 16 output channels. The propagation delay from the VME J2 connector to each slave slot is carefully equalized. Signal integrity measurements confirm a channel-to-channel timing skew of less than 100 ps, ensuring that the intrinsic timing precision of the FPGA is faithfully maintained at the physical output ports.

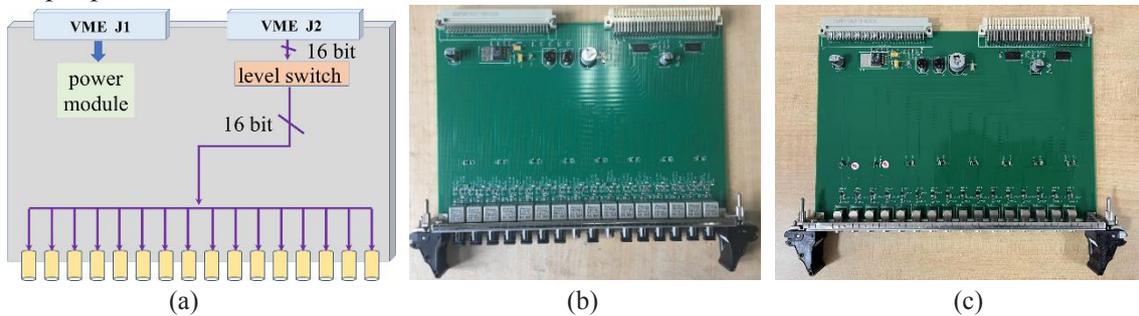

**Figure 4.** (a) Schematic of the output interface board. (b) Optical interface board. (c) Electrical interface boardThe functional design and implementation of the output interface board.

## 3.4 VME J2 Breakout Backplane Design

As depicted in the right panel of Fig. 5, a critical innovation in this design is the custom VME J2 breakout backplane. Taking advantage of the user-defined pins on the J2 connector of a standard VME crate, the system employs a dedicated J2 reversal backplane to establish point-to-point direct connections between the J2 custom pins of the master slot and those of the slave slots, as shown in figure 5(a). This backplane provides only physical-layer electrical interconnects and incorporates no active or passive electronic components, aside from the mechanical structures required for mating with VME high-speed connectors, as shown in figure 5(b). This minimalist approach significantly reduces parasitic capacitance, inductance, and propagation delays along the signal paths, thereby preserving signal integrity and maintaining precise timing synchronization across all channels. Consequently, it effectively mitigates the timing jitter and inter-channel skew typically associated with conventional shared-bus architectures. Figure 5(c) shows the rear view of the VME crate with the breakout backplane.

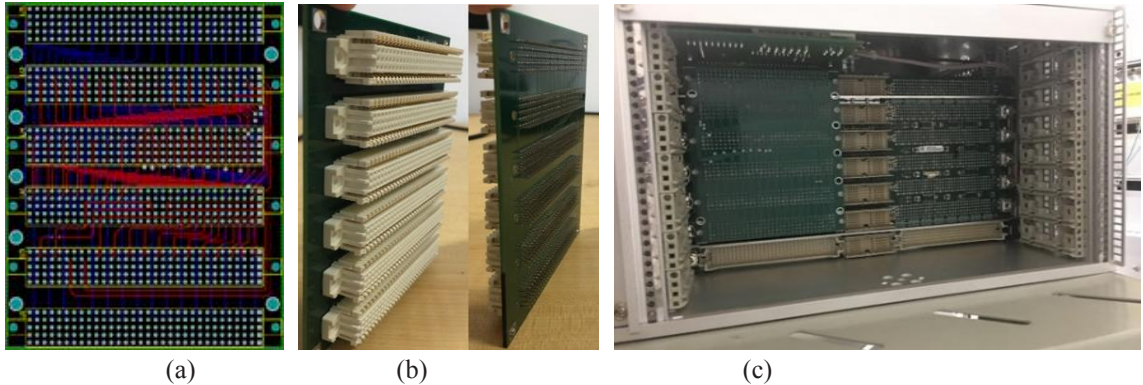

(a)  (b)  (c)

**Figure 5.** (a) PCB design (point-to-point). (b) VME J2 breakout backplane. (c) Rear view of the VME crate with the breakout backplane.

## 3.5 Remote Signal Terminal Board Design

As shown in figure 6(a), two types of signal termination units, housed in a compact 1U chassis, have been developed to address diverse application requirements:

- **standard opto-electronic conversion unit**: this unit employs a conventional opto-electronic circuit that converts one optical input into one electrical output. It accommodates two independent PCBs, each processing two optical inputs and delivering two corresponding electrical outputs, yielding a total of four opto-electronic channels per unit, as depicted in the right panel of figure 6(a).
- **high-fanout opto-electronic distribution unit**: designed primarily for beam diagnostics applications requiring multi-point synchronous triggering, this variant converts a single optical input into ten precisely synchronized electrical outputs. The design ensures minimal inter-channel timing jitter and excellent amplitude uniformity across all fanout paths, making it well-suited for time-critical beam instrumentation, as depicted in the right panel of figure 6(b).

The receiver employs an Intel MAX II CPLD (EPM1270). In contrast to FPGAs, which rely on segmented routing matrices and require strict physical constraints to maintain timing consistency across compilations, the CPLD's macrocell-based continuous routing matrix inherently guarantees highly predictable and deterministic pin-to-pin propagation delays ($t_{PD}$). Combined with its non-volatile configuration memory, this makes the CPLD an optimal and robust choice for such timing-critical terminal applications. Optical signals are received via HFBR-2412T photoreceivers, conditioned within the CPLD, and subsequently routed to local output drivers. A robust output stage has been implemented to translate the CPLD's 3.3 V logic levels into standard TTL-compatible signals with sufficient drive strength to reliably source 50 Ω terminated loads. Each terminal unit is equipped with both fiber-optic ports (using HFBR-1414T/2412T transceivers) for daisy-chaining or system expansion and LEMO connectors for local electrical outputs, thereby supporting precise synchronization across both remote and proximate instrumentation. Users may select the appropriate terminal variant based on specific experimental needs, enabling an optimal trade-off among functionality, channel density, and system cost.

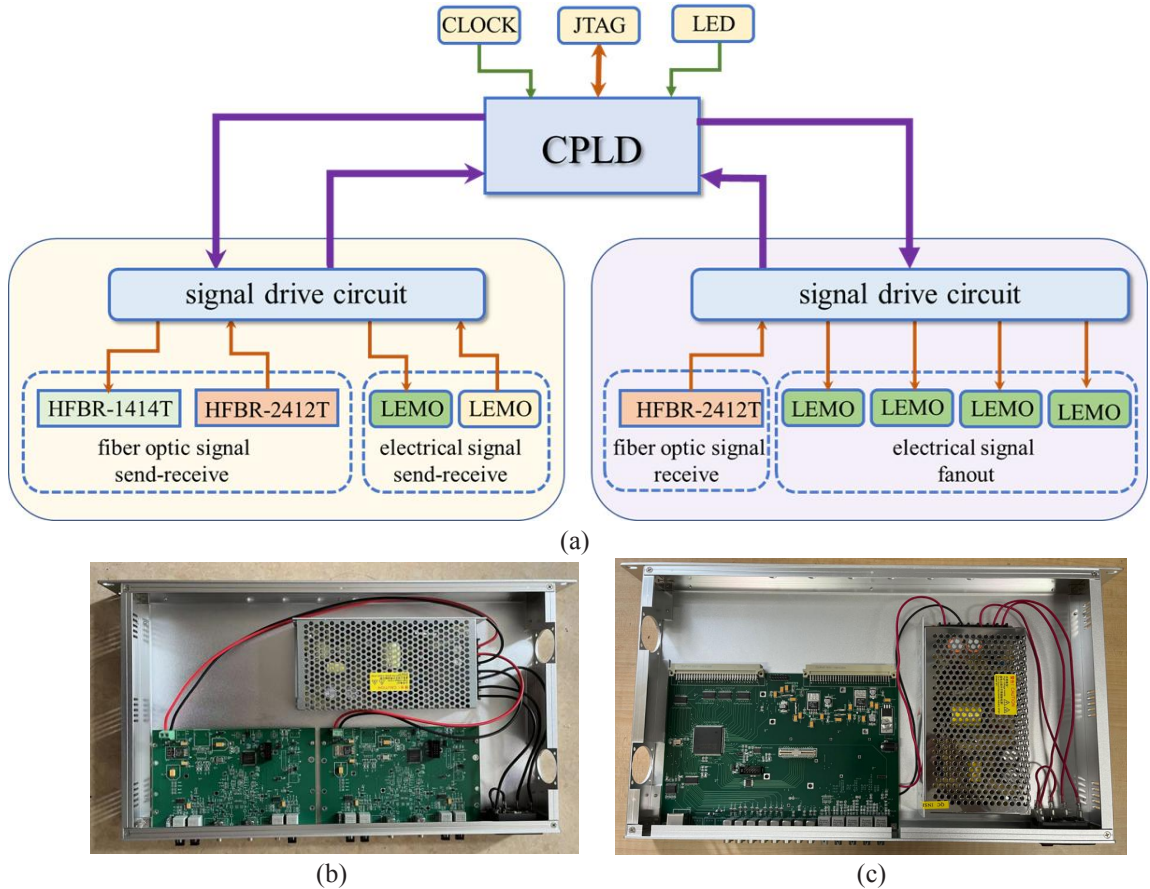

**Figure 6.** (a) Diagram of the Timing Interface Board. (b) Standard opto-electronic conversion unit. (c) High-fanout opto-electronic distribution unit.

## 4. System performance verification

To rigorously evaluate the timing precision and operational stability of the developed system, comprehensive bench tests were conducted using a high-bandwidth oscilloscope (Tektronix DPO7254C, 2.5 GHz bandwidth, 40 GS/s). The characterization focused on three critical metrics: synchronization jitter, pulse generation flexibility, and channel-to-channel synchronization accuracy.

As depicted in Figure 7(a), when synchronized to an external 648 MHz RF reference, the system demonstrated a peak-to-peak jitter of 164 ps and an RMS jitter of approximately 31.82 ps, measured between the RF zero-crossing and the output trigger edge. This confirms the FPGA's capacity for tight phase locking with the accelerator's RF reference. To strictly isolate and assess the intrinsic jitter of this timing distribution system, the focus was limited to the baseline electronic jitter generated solely by the master core logic board and the local electrical driver circuitry. Data was captured using the oscilloscope's infinite persistence mode, triggered explicitly by the internally synchronized master clock. Under these localized conditions, the electrical interface exhibited an ultra-low RMS jitter of 6.55 ps and a peak-to-peak jitter of 60 ps, as depicted in Figure 7(b). This metric quantifies the raw phase-locking precision and clock-generation limits of the master FPGA hardware itself. For remote distribution, jitter measurements were obtained at the output of the optical-to-electrical receiver following fiber transmission. Although fiber dispersion and conversion noise resulted in a peak-to-peak variation of 1000 ps, the RMS jitter remained at 119.5 ps, meeting the sub-nanosecond synchronization standard required for remote modulator triggering, and it is noted that the 10%–90% rise time of the TTL-compatible trigger signal is approximately 1 ns. Those performance are sufficient to maintain synchronization accuracy and drive the subsequent stage without introducing significant distortion. as depicted in Figure 7(c).

The system's functional versatility was demonstrated through pulse synthesis and timing adjustment tests. Figure 7(d) displays a continuous 100 Hz pulse train with a pulse width of 9.9 ms, illustrating the stability of the frequency synthesis logic and its ability to drive devices that require long gate signals. The fine-grained delay adjustment capability was further validated using the internal onboard oscillator. Figure 7(e) illustrates trigger output shifts in 6.18 ns increments, confirming the precision of the FPGA's high-speed counters.

Despite implementing strict isochronous routing on the output interface boards, residual timing skew persists across the 80 output channels. This skew arises mainly from minor trace-length mismatches on the core logic board and propagation delay variations introduced by the custom VME backplane interconnects. A dedicated calibration strategy was implemented to achieve sub-nanosecond synchronization accuracy and compensate for these hardware-induced discrepancies:

- Global Clocking: All output pins are managed by a single global clock distribution network (BUFG) to ensure uniform clock arrival times at the FPGA I/O banks [16].
- FPGA-based Fine Tuning: Precise temporal alignment is achieved using the native I/O delay resources of the Xilinx Spartan-6 FPGA [17]. Each output channel is managed using an ODDR (Output Double Data Rate) register followed by an IODELAY2 primitive configured in output delay (ODELAY) mode. The IODELAY2 provides programmable fine delay with a nominal tap resolution of approximately 200 ps, as shown in Figure 7(f).

Although the IODELAY2 primitive exhibits inherent tap-to-tap nonlinearity across process, voltage, and temperature (PVT) variations, this non-linearity is practically negligible for this application. During initial system commissioning, a meticulous channel-by-channel calibration is performed: the channel with the maximum intrinsic propagation delay is selected as the baseline, and the remaining 79 channels are statically tuned using the IODELAY2 taps. Because the system's global requirement is sub-nanosecond (< 1 ns) synchronization alignment, an average single tap resolution of ~200 ps provides adequate granularity to compensate for static backplane skew, fully absorbing any PVT-induced step variations. Based on pre-operational calibrations, individual tap values are calculated and loaded into each IODELAY2 primitive during initialization. This alignment ensures that all 80 outputs conform to a common timing reference within a single tap step, maintaining the total signal path skew, across the custom backplane and output interface boards, well below the 1 ns design specification.

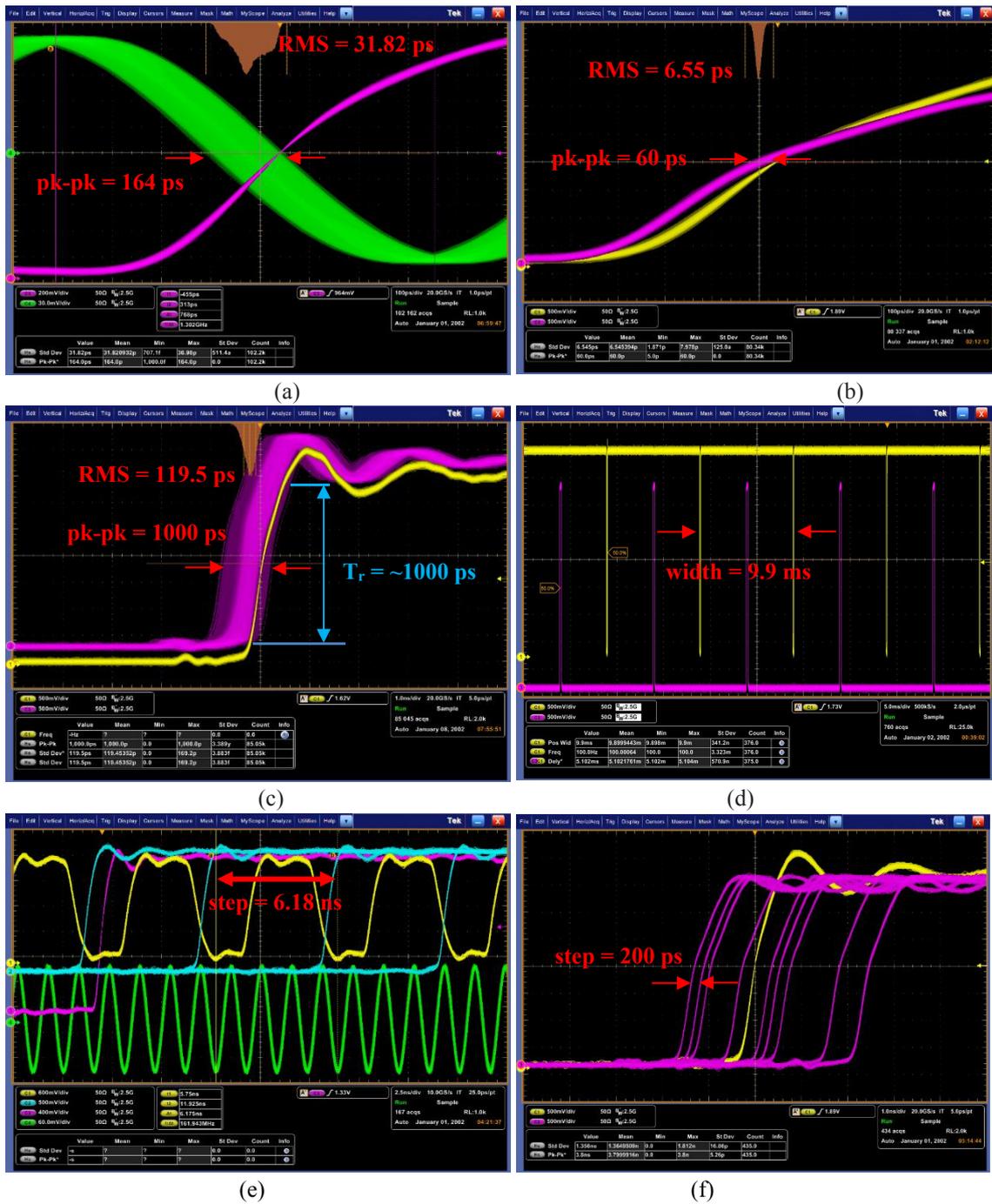

**Figure 7.** (a) RMS jitter (32.12 ps) of the electrical trigger synchronized to an external 648 MHz RF reference. (b) Verification of pulse generation showing a stable 100 Hz output. (c) RMS jitter (6.55 ps) of the local electrical output trigger and the achieved 10% to 90% rise time of the TTL-compatible trigger signal is approximately 1 ns. (d) RMS jitter (119.5 ps) of the optical trigger after fiber transmission. (e) Delay adjustment steps (6.18 ns) utilizing the external RF reference. (f) Delay adjustment steps (~ 200 ps) utilizing the delay line of IODELAY2.

## 5. Scalability Design

A highly stable and precise clock-and-reset synchronization mechanism enables the expansion of synchronized output channels to 160, fulfilling the dual demands of channel count and sub-nanosecond timing accuracy required by medium-scale facility in high-performance timing distribution systems. To minimize the number of interface board variants and enable bidirectional synchronization control, the two core logic boards exchange critical timing signals through integrated small form-factor pluggable (SFP) optical transceivers[18]. As depicted in figure 8(a), the core logic board B transmits a 100 MHz reference clock over fiber to the core logic board A, while the board A returns a reset signal via a separate fiber link to the board B. This bidirectional closed-loop configuration ensures strict temporal alignment between the timing signals generated by both boards. Further, the FPGA's general-purpose differential I/O banks, configured for LVDS/LVPECL-compatible signaling, are directly used to drive the SFP optical transceivers. Dedicated level-shifting ICs are inserted between the FPGA outputs and the SFP modules, as depicted in figure 8(b). The entire signal path is optimized through precise 50 Ω impedance matching and meticulous PCB layout practices, including controlled trace lengths and minimized stubs.

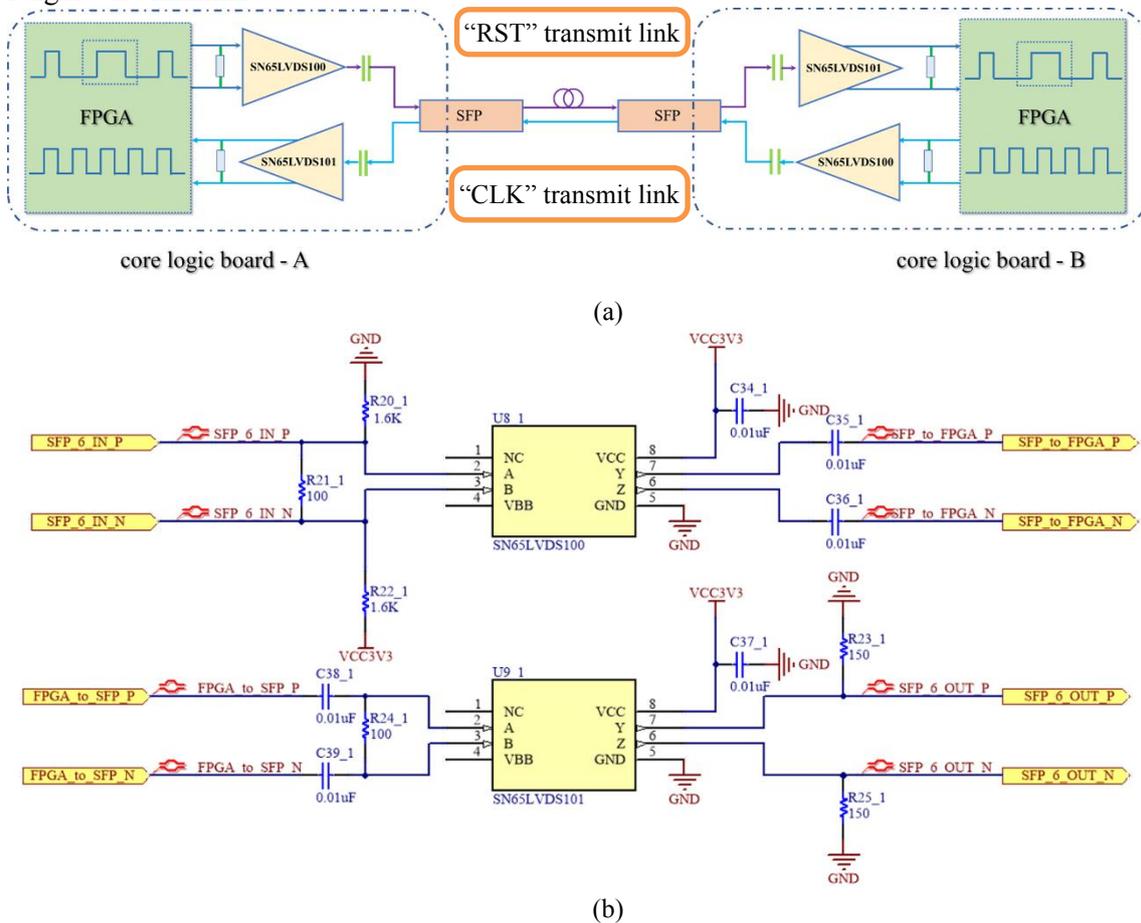

(a)

(b)

**Figure 8.** (a) Schematic diagram of the direct transmission between two CLB via SFP connectors. (b) Schematic design of the signal transmission level-shifting circuitry.

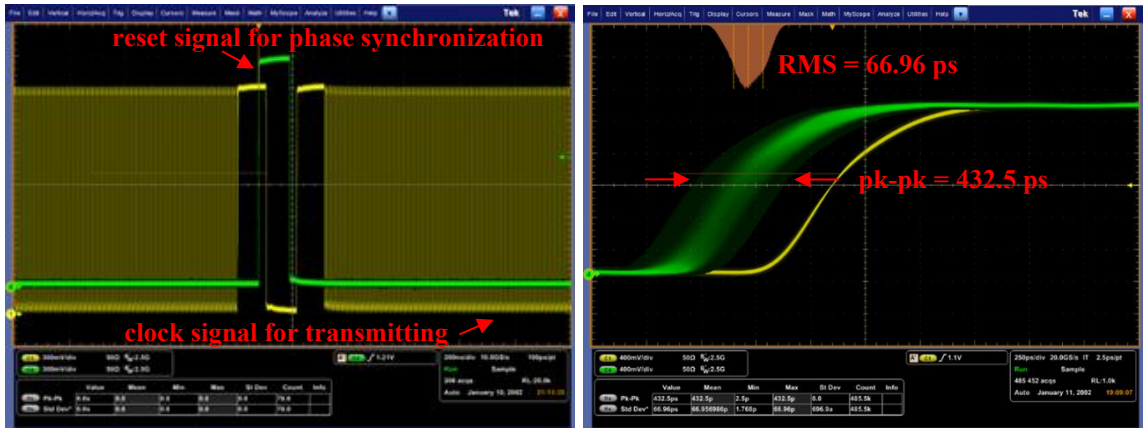

(a) Reset pulse embedded into a continuous 100 MHz clock stream    (b) Jitter of the reset pulse
**Figure 9.** Test of the reset pulse using the direct transmission between two CLB via SFP connectors.

While advanced commercial systems (e.g., White Rabbit and MRF) effectively utilize the Clock-Data-Recovery (CDR) circuits within GTP/GTX transceivers, often combined with precise phase-tracking mechanisms like DDMTD, to achieve ultra-low jitter (<50 ps), a standard logical instantiation of the GTP protocol can be problematic for deterministic timing [19]. Specifically, without complex deterministic latency modifications, the standard GTP RX elastic buffer and comma alignment logic introduce a phase ambiguity of up to one parallel clock cycle upon reset or power-up. To circumvent the need for developing complex elastic-buffer-bypass and CDR phase-tracking logic on the cost-effective Spartan-6 architecture, this system employs a simplified, custom encoding scheme. To overcome this limitation, the reset pulse is encoded within the FPGA using dedicated logic resources. The pulse is embedded into a continuous 100 MHz clock stream as a quasi-balanced code that satisfies the DC-balance and transition-density constraints of the optical module, as depicted in figure 9(a). At the receiver, the timing edge is accurately reconstructed through synchronous decoding, enabling lossless recovery of the pulse with sub-nanosecond fidelity. This design strategy eliminates clock-domain-induced jitter, significantly improves signal integrity, and reduces propagation delay variation. As a result, an RMS jitter of 66.96 ps (as depicted in Figure 9(b)) is consistently achieved for the direct point-to-point synchronization link between two CLBs via SFP transceivers. Furthermore, when deployed in the complete macroscopic distribution chain—routing from the master CLB through the custom backplane and optical interface boards, transmitting across long-distance fiber, and finally undergoing optoelectronic conversion at the remote CPLD terminal—the ultimate triggering output maintains a highly robust sub-nanosecond precision (RMS jitter = 119.5 ps, as shown in Figure 7(c)), fully satisfying the system's demanding requirements for remote modulator triggering.

## 6. System Integration and Operational Deployment

The fully developed timing distribution system has been successfully commissioned and integrated into a standard 19-inch instrumentation rack at the C-band photocathode electron gun test platform, as illustrated in figure 10. A commercial RF signal generator functions as the facility's master oscillator, providing a stable reference RF signal to the Core Logic Board via its front-panel input. The cabling layout demonstrates the system's high-density fan-out capability and hybrid transmission strategy. Orange fiber-optic cables are connected to the optical interface boards to distribute synchronized triggers to remote subsystems, such as high-voltage modulators and the drive laser system. This optical transmission ensures complete galvanic isolation and robust immunity to the electromagnetic interference generated by high-power klystrons. Concurrently, electrical trigger signals are routed via coaxial cables to local beam diagnostics and monitoring equipment, facilitating convenient local access. The control architecture is built upon the Experimental Physics and Industrial Control System (EPICS) framework. Communication between the host IOC and the FPGA logic is established over a serial interface utilizing standard EPICS support modules: streamDevice for protocol-level communication, and asynDriver for asynchronous port management. Low-level hardware registers are mapped to EPICS Process Variables (PVs) using standard analog input (ai) and analog output (ao) records, enabling read and write access to timing parameters. Operator interaction is mediated through an Operator Interface (OPI) developed within the Control System Studio (CSS) environment. As shown in figure 11, the OPI provides a centralized dashboard for the management of all 80 trigger channels. Key operational parameters, such as repetition rate, delay, pulse width, pulse count, and gating modes, can be remotely configured in real-time. Furthermore, the interface incorporates status readback mechanisms and configurable alarm limits, significantly enhancing operational safety and efficiency during high-voltage conditioning and beam commissioning.

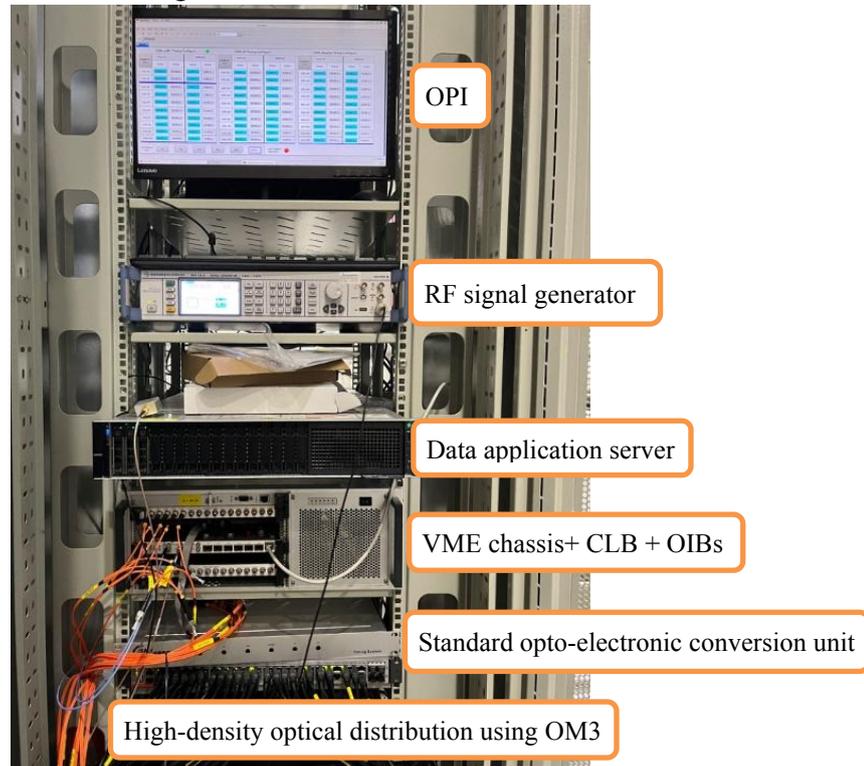

**Figure 10.** Photograph of the timing system deployed at the C-band photocathode gun test platform, including: (Top) The EPICS/CSS-based OPI for remote configuration; (Middle) An RF signal generator providing the master reference clock; (Bottom) The 6U VME chassis housing the core logic board (center) and multiple output interface boards. Orange fiber cables demonstrate the high-density optical distribution to remote accelerator subsystems.

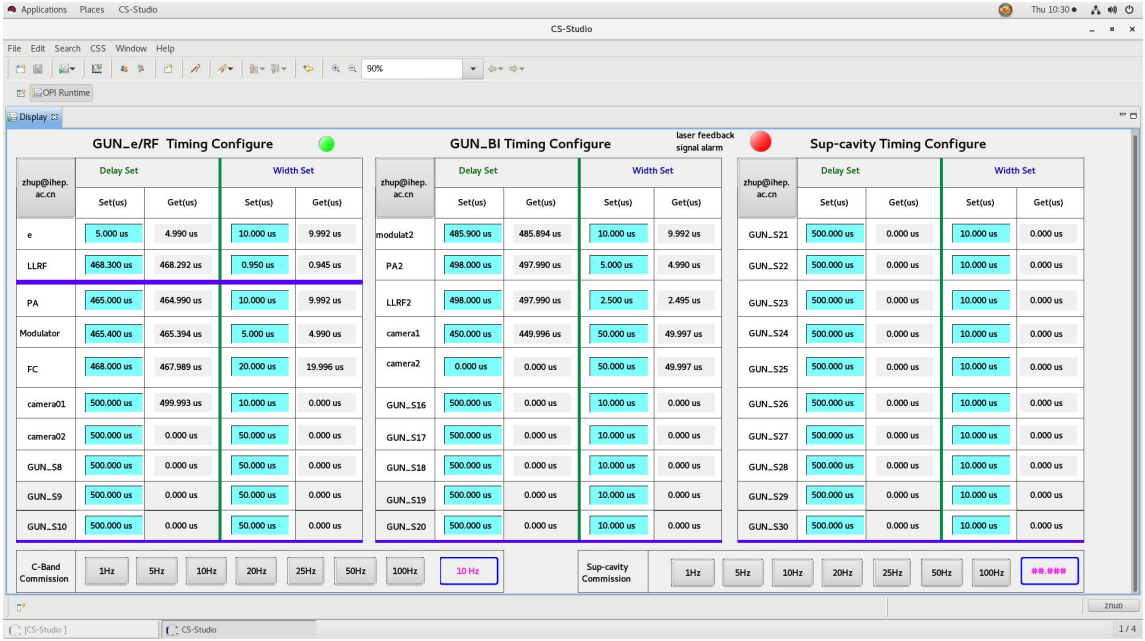

**Figure 11.** The timing OPI developed in CSS/EPICS. The panel allows for individual configuration of delay and width for Gun RF, Laser, and diagnostic triggers, providing real-time status monitoring.

The operational timing logic was rigorously optimized in coordination with the high-power RF and laser subsystems. Crucially, figure 12(a) demonstrates the phase alignment between the photocathode drive laser and the RF field. The oscilloscope trace confirms that the laser pulse (Yellow) is strictly synchronized within the accelerating phase of the RF envelope (Blue). This precise temporal overlap ensures efficient photoelectric emission and optimal acceleration. The observed stability of this synchronization confirms the system's capability to maintain sub-nanosecond jitter performance in a high-noise environment.

To experimentally validate the overall performance of the timing distribution system, on-line beam commissioning was conducted at the C-band photocathode gun test platform. The facility operated at a repetition rate of 10 Hz, with a beam energy of 5 MeV and a bunch charge of 100 pC. Under these conditions, the transverse emittance was characterized using the solenoid scan technique. Figure 12(b) presents the transverse beam profile captured on the YAG screen. The high shot-to-shot stability of the beam spot size and centroid position indicates that the timing system effectively minimizes jitter-induced instabilities in the laser arrival time relative to the RF phase. The successful transport and characterization of the 100 pC electron bunches verify that the developed timing system meets the stringent synchronization requirements necessary for high-brightness C-band photocathode gun test platform.

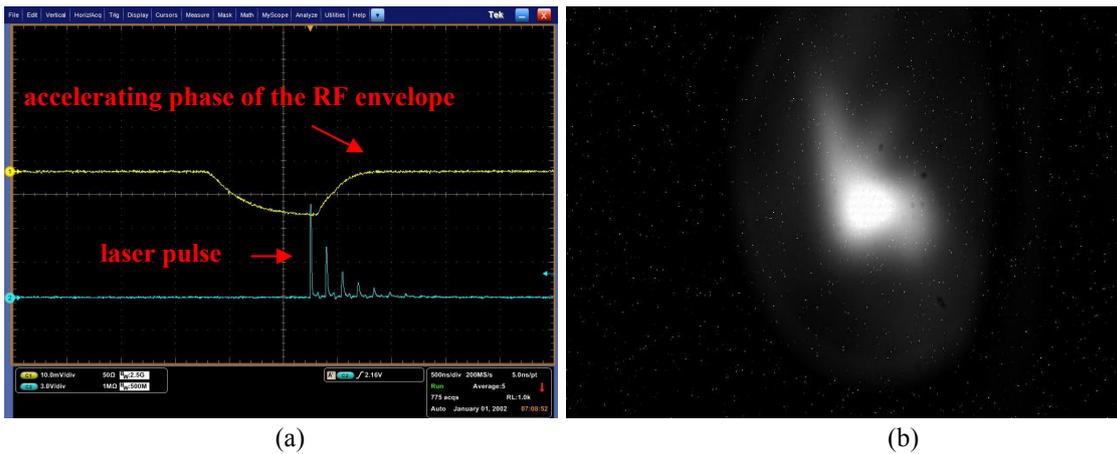

(a)                          (b)

**Figure 12.** (a) Verification of the synchronization between the photocathode drive laser and the RF field. (b) Transverse beam profile measured on a YAG screen.

## 7. Summary

This paper presents the design, implementation, and validation of a high-density, cost-effective trigger distribution system tailored for the C-band photocathode gun test platform. By innovatively repurposing the standard 6U VME architecture with a custom point-to-point backplane interconnect, the system integrates an FPGA-based master controller within a star topology. This configuration ensures deterministic signal propagation across up to 80 independent channels. Additionally, the modular design, featuring interchangeable optical and electrical interface boards, provides exceptional adaptability for diverse experimental setups, while the "serial server + virtual machine IOC" framework facilitates real-time, intuitive remote configuration.

    Experimental characterization confirms that the system meets stringent high-precision synchronization requirements. A dual-clocking capability ensures seamless synchronization with both local timebases and facility-wide RF references. The system supports adjustable trigger frequencies from 1 Hz to 100 Hz, with delay and pulse widths tunable over a 0–10 ms range at a resolution of 10 ns (or the RF period). In terms of signal stability, the local electrical output exhibits an RMS jitter of less than 6.55 ps. For remote optical distribution, the system maintains a sub-nanosecond RMS jitter of 119.5 ps, with peak-to-peak drift limited at 1000 ps due to fiber transmission effects. Even after optical transmission and electrical conversion, the RMS jitter remains well within the sub-nanosecond range. Furthermore, by employing a compact dual-channel SFP optical signaling architecture, the system offers seamless scalability to 160 synchronized output channels. This capability significantly enhances channel density and deployment efficiency while maintaining sub-nanosecond timing accuracy, thereby fully satisfying the rigorous requirements of medium-scale particle accelerators for both the quantity and precision of timing signals.

1
2

## Acknowledgments

This work was supported by the funds of National Natural Science Foundation of China (Fund No: 11875270, 12205317). The authors would like to sincerely thank the C-band photocathode gun test platform builders and contributors for their hard work and dedication.